\documentclass[useAMS,usenatbib]{mn2e}
\usepackage{longtable,color}
\usepackage{graphicx}
\usepackage{mathrsfs}
\usepackage{epsfig}
\usepackage{natbib}
\usepackage{color}
\usepackage{float}
\usepackage{amsmath}
\usepackage{times}
\usepackage{upgreek}
\usepackage[varg]{txfonts}
\usepackage{enumerate}
\usepackage{verbatim}
\usepackage{booktabs}
\usepackage{multirow}
\usepackage{amssymb}
\usepackage{placeins}

\title[UDGs in MS2]{A Universe of Ultra-Diffuse Galaxies: Theoretical
  Predictions from $\Lambda$CDM Simulations}

\author[Rong et al.]{Yu Rong$^{1}$\thanks{E-mail: rongyu@nao.cas.cn},
  Qi Guo$^{1}$, Liang Gao$^{1,2}$, Shihong Liao$^{1}$, Lizhi
  Xie$^{3}$, Thomas H. Puzia$^{4}$, \and Shuangpeng Sun$^{1}$, Jun
  Pan$^{1}$\\
    $^{1}$Key Laboratory for Computational Astrophysics, The Partner Group
  of Max Planck Institute for Astrophysics, National Astronomical
  Observatories,\\ Chinese Academy of Sciences, Beijing 100012,
  China\\
  $^{2}$Institute of Computational Cosmology, Department of Physics,
  University of Durham, Science Laboratories, South Road, Durham DH1
  3LE, UK\\
  $^{3}$OATS, INAF, Via Bazzoni 2, 34124-Trieste, TS, Italy\\
  $^{4}$Instituto de Astrof\'isica, Pontificia Universidad Cat\'olica
  de Chile, Av. Vicu\~na Mackenna 4860, 7820436 Macul, Santiago, Chile
  }

\begin{document}
\maketitle

\begin{abstract}
A particular population of galaxies have drawn much interest recently,
which are as faint as typical dwarf galaxies but have the sizes as
large as $L^*$ galaxies, the so called “ultra-diffuse galaxies”
(UDGs). The lack of tidal features of UDGs in dense environments
suggests that their host halos are perhaps as massive as that of
the Milky Way. On the other hand, galaxy formation efficiency should
be much higher in the halos of such masses. Here we use the model galaxy
catalog generated by populating two large simulations: the
Millennium-\uppercase\expandafter{\romannumeral2} cosmological
simulation and Phoenix simulations of 9 big clusters with the
semi-analytic galaxy formation model. This model reproduces remarkably
well the observed properties of UDGs in the nearby clusters, including the
abundance, profile, color, and morphology, etc. We search for UDG
candidates using the public data and find 2 UDG candidates in our
Local Group and 23 in our Local Volume, in excellent agreement with
the model predictions. We demonstrate that UDGs are genuine dwarf
galaxies, formed in the halos of $\sim 10^{10}M_{\odot}$. It is the
combination of the late formation time and high-spins of the host
halos that results in the spatially extended feature of this
particular population. The lack of tidal disruption features of UDGs
in clusters can also be explained by their late infall-time.

\end{abstract}
\section{Introduction}

A population of low surface brightness galaxies has been observed in
spatial regions of, e.g., Coma
\citep{vanDokkum15a,vanDokkum15b,Koda15,Yagi16}, Virgo
\citep{Mihos15}, Fornax \citep{Munoz15}, A168 \citep{Roman16a}, A2744
\citep{Janssens17}, eight other clusters with redshifts $z\sim
0.044\-- 0.063$ \citep{vanderBurg16}, and Pisces-Perseus supercluster
\citep{Martinez-Delgado16}, as well as several galaxy groups
\citep{Makarov15,Toloba16,Roman16b,Trujillo17}. While their stellar
masses are similar to typical dwarf galaxies, their effective radii 
are similar to the $L^*$ galaxies \citep{vanDokkum15a,Beasley16}. These galaxies are generally referred to as 
ultra-diffuse galaxies (UDGs). Except for several blue ones
\citep[e.g.,][]{Roman16a}, the majority of the observed UDGs populate
the red sequence, suggesting that the star formation in UDGs has been
quenched before $z\sim 2$ \citep{vanderBurg16}.

UDGs are ubiquitously distributed from the cores of galaxy clusters to
the surrounding large-scale filaments
\citep{Koda15,Yagi16,Roman16a,Roman16b}. Since they can withstand the
strong tidal forces in the cluster cores without significant features
of tidal disruption, one scenario is that UDGs are dark matter
dominated galaxies, for instance, the failed $L_\star$ galaxies which
lost their gas content after the first generation of stars (van Dokkum
et al. 2015a; Scannapieco et al. 2008; Stinson et al. 2013;
Trujillo-Gomez et al. 2015). Using stellar kinematics of Dragonfly 44,
van Dokkum et al. (2016) measured its dynamical mass as $\sim
10^{12}\ M_{\odot}$, similar to the mass of the Milky Way; this is
unexpected from prediction of subhalo abundance matching (SHAM) in which galaxy formation efficiency reaches its maximum at this halo mass \citep{Guo10,Conroy09,Simha12}. Another scenario is that UDGs are spatially
extended dwarf galaxies \citep[e.g.,][]{Amorisco16,Cintio16,Dalcanton97a,Dalcanton97b,Mo98,Huang12}. Using
the abundance and kinematics of globular clusters around two UDGs,
VCC~1287 and DF17, \cite{Beasley16} and \cite{BeasleyTrujillo16}
estimated the corresponding dynamical masses to be around
$m_{\rm{vir}}\sim (8\pm 4)\times 10^{10}\ M_{\odot}$ and
$m_{\rm{vir}}\sim (9\pm 2)\times 10^{10}\ M_{\odot}$ respectively,
similar to that of the typical dwarf galaxies. Using the relation
between the mass of the globular cluster system and the halo mass
\citep{Harris13,Harris15}, \cite{Peng16} also inferred the total mass of DF17
to be $(9.3\pm 4.7)\times 10^{10}\ M_{\odot}$; \cite{Amorisco16b}
estimated the dynamical masses of 54 Coma UDGs to be lower than
$1.3\times 10^{11}\ M_{\odot}$. \cite{Roman16a} also found that the
distribution of UDGs around A168 is similar to the normal dwarfs, but
significantly different from the distribution of massive galaxies with
masses similar to the Milky Way. Using the width of the HI line, Trujillo
et al. (2016) found a UDG in the very local Universe with a virial mass
of 8 $\times 10^{10}M_{\odot}$. Theoretical work also suggest that
UDGs might be genuine dwarf galaxies possibly with high spins
(Amorisco \& Loeb 2016, Yozin \& Bekki 2015), or spatially extended
stellar components caused by feedback driven gas outflows
\citep{Cintio16}.

In this paper, we will use a publicly available semi-analytic galaxy
catalog \citep{Guo13} to investigate whether UDGs can naturally emerge
from the $\Lambda$CDM hierarchical structure formation model. In
section 2, we briefly describe the simulation and the
semi-analytic models, as well as the selection criteria of the
possible UDGs. In section 3, we compare the model predictions with
observational results. In section 4, we study the distributions of the model UDGs, and dependence of UDG properties on
environments. In section 5, we investigate the origin of this
particular population. Conclusions are presented in section 6.

\section{UDG selection from simulations}

\subsection{Simulations}

The galaxy catalogs used here are based on two simulations,
the Millennium-\uppercase\expandafter{\romannumeral2} simulation
(MS-\uppercase\expandafter{\romannumeral2}; Boylan-Kolchin et
al. 2009), and Phoenix simulation (Gao et
al. 2012). MS-\uppercase\expandafter{\romannumeral2} is a
high-resolution cosmological $N$-body simulations, following $2160^3$
particles from $z=127$ to $z=0$ in a periodic box of
$100\ {\rm{Mpc}}/h$ on a side. Each dark matter particle has a
mass of $6.88\times 10^6\ M_{\odot}/h$. Particle data were
stored at 68 logarithmically spaced output times. The
MS-\uppercase\expandafter{\romannumeral2} simulation adopted the
cosmological parameters consistent with the first-year {\it {Wilkinson
    Microwave Anisotropy Probe}} (WMAP) result; it was then rescaled
to that consistent with WMAP seven-year parameters \citep{Guo13}:
$\Omega_{\rm{m}}=0.272$, $\Omega_{\rm{b}}=0.0455$, $h=0.704$,
$\sigma_8=0.81$, $n=0.967$. The Phoenix simulation is a high-resolution
re-simulation of nine individual rich clusters and their
surroundings. Each Phoenix cluster has been simulated at different
resolution levels for numerical convergence studies. Here we adopt the
simulation with level 2 resolution. Particle mass is of $\sim
10^6\ M_{\odot}/h$, which varies from cluster to cluster
slightly. The Phoenix simulation adopted cosmological parameters from
a combination of Two-degree-Field Galaxy Redshift Survey (Colless et
al. 2001) and first-year Wilkinson Microwave Anisotropy Probe data
(Spergel et al. 2003):  $\Omega_{\rm{m}}=0.25$,
$\Omega_{\rm{b}}=0.045$, $h=0.73$, $\sigma_8=0.9$, $n=1$. Although the
cosmological parameters adopted by these two simulations are slightly
different, this has negligible effect on our main results.

At each snapshot, dark matter halos are identified with the
friends-of-friends (FOF) group algorithm by linking particles
separated by 0.2 times the mean inter-particle separation (Davis et
al. 1985). The SUBFIND algorithm (Springel et al. 2001) was then used to
identify self-bound subhalos; merger trees were constructed by linking
each subhalo at different output times to its unique descendant using
the algorithm described in Springel et al. (2005) and Boylan-Kolchin
et al. (2009).

MS-\uppercase\expandafter{\romannumeral2} contains millions of halos
from $ 10^{10}\ M_{\odot}/h$ to $10^{14}\ M_{\odot}/h$,
allowing us to study the possible UDGs in different environments in a
statistical way. Yet limited by the box size,
MS-\uppercase\expandafter{\romannumeral2} has no clusters as massive
as the Coma clusters, $\sim 2\times 10^{15}\ M_{\odot}$, and A2744,
$\sim 5\times 10^{15}\ M_{\odot}$, where the largest samples of UDGs
are discovered. The Phoenix simulation suits compensate this by
providing more massive cluster samples, and the largest cluster in the
Phoenix suits has a mass $\sim 3.4\times 10^{15}\ M_{\odot}$
(at $z=0$). The minimum resolved halo in this Phoenix cluster is of mass
$3.6\times 10^{8}\ M_{\odot}/h$, well below the mass limit
($\sim10^9M_{\odot}$) below which haloes can not form any galaxies. We
thus use the combination of these two simulation sets to study UDGs in
various environments.

\subsection{Galaxy formation model}
In order to populate dark matter halos with galaxies, we applied the
semi-analytic galaxy formation models (Guo et al. 2011, 2013) to the
stored subhalos merger trees extracted from these $N$-body
simulations. This model has been proved successful in reproducing many
galaxies properties both in the local Universe and at high redshifts,
and particularly it provides convincing results for galaxy size
vs. stellar mass relations. Here we briefly summarize the main
physical processes relevant to the formation of galaxies as faint as
UDGs and the models of galaxy stellar component sizes.

As discussed in Guo et al. (2011; here after Guo11), two processes are
crucial for the formation of low mass galaxies: UV reionization and
supernova (SN) feedback. The capability to capture baryons is reduced
in low mass systems due to the UV reionization. \cite{Guo11,Guo13}
adopted results given by \cite{Okamoto08} to quantify the fraction of
baryons as a function of halo mass. As demonstrated in Guo11
this effect becomes significant for galaxies less luminous than $M_V
=$ -11. Vast amount of energy is released during SN explosion which can reheat the
surrounding gas and even eject gas out of its dark halo. Guo11 introduced a SN feedback model which depends on the maximum
velocity of the host halo, leading to a relatively more efficient
feedback in low mass halos than their high mass counterparts. This
significantly changes the slope of the stellar mass function at the
low mass end (e.g. stellar mass $m_{\rm{st}} < 10^{9.5}\ M_{\odot}$). 

Guo11 uses the stellar population synthesis models from Bruzual \& Charlot (2003), and adopts a Chabrier initial function to calculate the photometric properties of galaxies. For low-redshift galaxies, the slab dust model introduced in De Lucia \& Blaizot (2007) is then implemented to account for the extinction of star light. Further comparison with observations suggests that the fiducial model of Guo11 can well predict the luminosity function of galaxies at low redshifts \citep[e.g.,][]{Guo11,Nierenberg12}, particularly in the faint end.

Our model galaxies contain two components, disks and bulges. Guo11 assumed
the stellar disk to have an exponential surface density profile. Its
size is determined by the specific angular momentum and the circular
velocity (here using $V_{\rm{max}}$ as a proxy). The angular momentum is
obtained from its gas disk during the star formation. The gas disk
acquires its angular momentum during the cooling process, i.e. the
cooling gas is assumed to have the same specific angular momentum as
its host halo. Bulges are formed by mergers and disk instability. In
Guo11, bulge sizes were calculated by assuming energy conservation and
virial equilibrium. For mergers, the relevant components are the
binding energy and interaction energy of the two merging galaxies. For
disk instability, they are the binding energy and interaction energy
of the existing bulges and the part of mass which is transferred into
bulges during disk instability.

In order to compare with the observations directly,
we convert the 3-D radius of our model galaxies to
effective radius $r_{\rm{e}}$ (also referred to as the projected
half-mass radius)  by assuming the stellar bulge and stellar disk to
have the Jaffe \citep{Jaffe83} and exponential density profiles,
respectively (see Xie et al. 2015 for details). For each galaxy, we
divide its projected radius (from $10\ \rm{pc}$ to $10\ \rm{kpc}$)
into 100 bins ($r_i$, where $i$ is from 1 to 100) in logarithm scale
and calculate the projected absolute magnitude $M_{i}^{\rm{pro}}$ of
each bin. The central surface brightness $\mu_0$ is obtained by
fitting the surface brightness $\mu_{i}\sim
M_{i}^{\rm{pro}}+2.5~{\rm{log}}_{10}(s_i)$, where $s_i$ is the area of
the bin, using a S\'ersic model. We test the dependence of our results on bin sizes and find
that this effect is minor.

\subsection{Sample selection}

\begin{figure}
\centering
\includegraphics[scale=0.29]{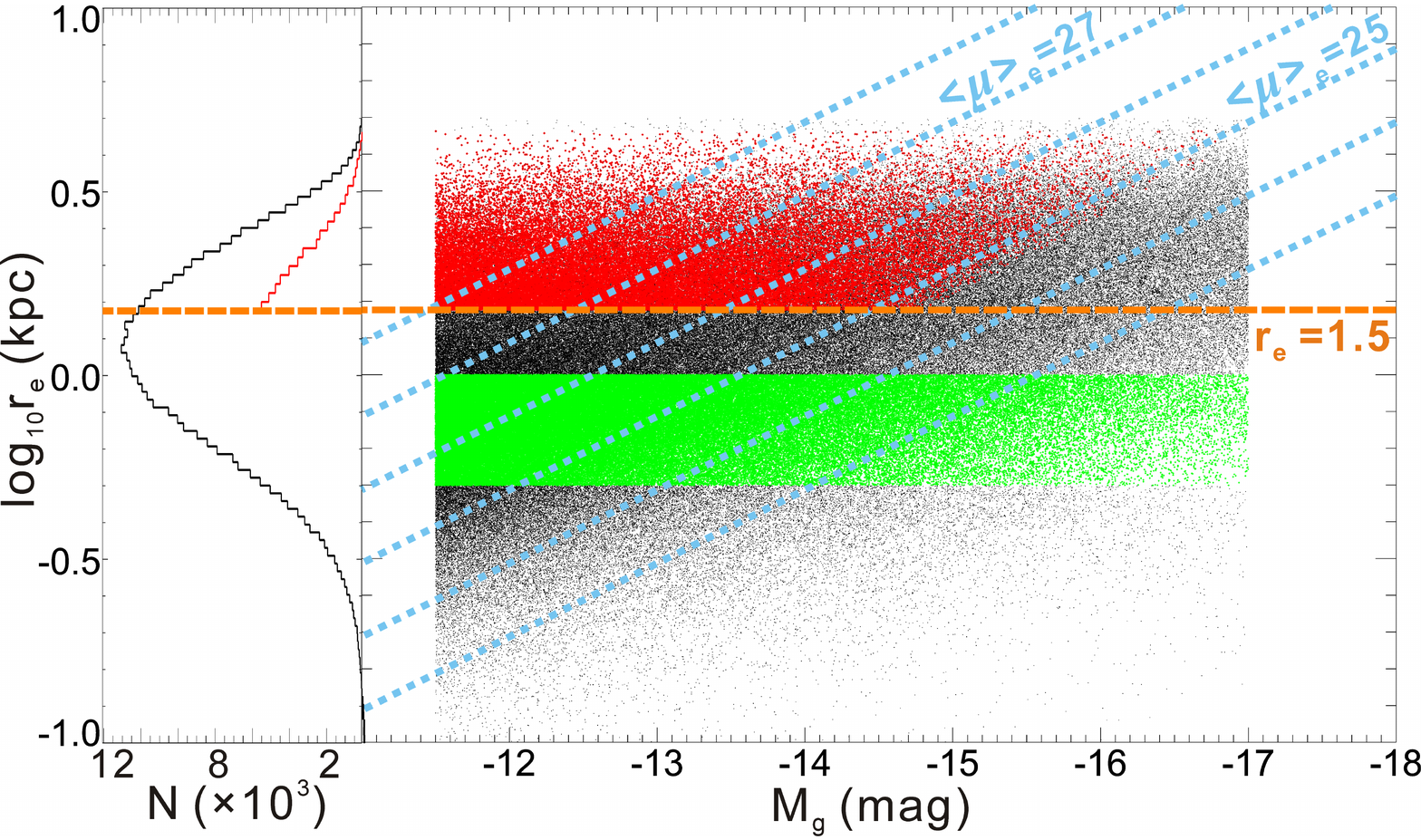}
\caption{Left panel: $r_{\rm{e}}$ distribution for model dwarf galaxies (black histogram) and UDGs (red histogram). Right panel: $r_{\rm{e}}\-- M_g$ relation for model galaxies. Model UDGs with $r_{\rm{e}}\in (1.5, 4.6)\ \rm{kpc}$, $M_g\in (-17, -11.5)\ \rm{mag}$, and $\mu_0 >23.5\ \rm{mag/arcsec^2}$ are denoted by the red points. Green points mark the dwarf counterparts with $r_{\rm{e}}\in (0.1, 1.5)\ \rm{kpc}$, $M_g\in (-17, -11.5)\ \rm{mag}$. Blue tilted dotted lines show the different mean surface brightness thresholds, and the orange horizontal dashed line denotes the threshold of $r_{\rm{e}}=1.5\ \rm{kpc}$.}
\label{re_mag}
\end{figure}

There are a few ways to define UDGs in the
literature. \cite{vanDokkum15a,vanDokkum15b} defined UDGs as the
galaxies with $\mu_{0,B}>24\ \rm{mag/arcsec^2}$ and
$r_{\rm{e}}>1.5$~kpc; some others used a slightly different quantity,
the mean surface brightness within $r_{\rm{e}}$,
$\langle\mu\rangle_{{\rm{e}},r}\geq 24\ \rm{mag/arcsec^2}$
\citep[e.g.][]{Yagi16,vanderBurg16,Janssens17}. Given that the
S\'ersic indices, $n$, of most UDGs are around 1
\citep{Yagi16,Koda15,Roman16a} and colors $g-r\sim 0.6$
\citep{vanderBurg16}, $\langle\mu\rangle_{{\rm{e}},r}\geq 24\ \rm{mag/arcsec^2}$ is
approximately equivalent to $g$-band
$\mu_{0,g}>23.5\ \rm{mag/arcsec^2}$ \citep{Graham05}. In this work, we
adopt the criteria as follows

\begin{itemize}
\item[] $1.5 < r_{\rm{e}} < 4.6\ {\rm{kpc}}$,
\item[] -17 $< M_g  <$-11.5~mag,
\item[] and $\mu_{0,g} > 23.5\ \rm{mag/arcsec^2}$,
\end{itemize}
where $M_g$ is the $g$-band absolute magnitude.

We show the $r_{\rm{e}}\-- M_g$ relation for the model galaxies in
Fig.~\ref{re_mag}. Different mean surface brightness
$\langle\mu\rangle_{\rm{e}}$ thresholds are highlighted with the dotted
lines. We find that almost all of the UDG candidates (red dots) are
distributed above $\langle\mu\rangle_{\rm{e}}=25\ \rm{mag/arcsec^2}$.
The luminosities of the observed UDGs are similar to those of the
typical dwarf galaxies; however, their sizes are much
larger. Interestingly, Fig.~\ref{re_mag} shows that UDGs are not an
isolated population, rather they exist as a continuous extension of
typical dwarf galaxies. For galaxies of similar magnitudes, the UDGs
occupy the large-size tail of the size distribution, suggesting that
UDGs are indeed a subsample of dwarf galaxies.

In order to understand the properties of UDGs and compare them with
the typical dwarf galaxies more clearly, we select a counterpart
sample of the dwarfs within the same luminosity range as the UDGs, but
different sizes of $r_{\rm{e}}\in (0.5, 1.0)\ \rm{kpc}$
\citep{vanderBurg16,Misgeld11}. According to the studies of
\cite{Graham03} and \cite{Mo10}, these dwarf counterparts primarily
include the dwarf ellipticals (dE) and dwarf spheroidals (dSph). Note
that the ultra-compact dwarfs are not included in this sample. In
total, we have $4.4\times 10^4$ UDGs and $1.3\times 10^5$ dwarf
counterparts, corresponding to 11\% and 32\% of the faint galaxies
($M_g \sim $-17 $\--$ -11.5) respectively, at $z=0$ in
MS-\uppercase\expandafter{\romannumeral2}.

\section{UDGs in simulations and observations}
In this section we will firstly compare the properties of the model
UDGs with the observations, and then explore how they vary with the
different environments.

\subsection{UDGs in clusters}
Most of the observed UDGs are discovered in clusters, e.g., in Coma
(van Dokkum et al. 2015a), A168 \citep{Roman16a}, A2744
\citep{Janssens17}, and other 8 low-redshift clusters
\citep[vdB16;][]{vanderBurg16}.  Here we focus on the comparison of
the UDGs in clusters between the model predictions and observations.

{\bf Abundance} 

\begin{figure}
\centering
\includegraphics[scale=0.35]{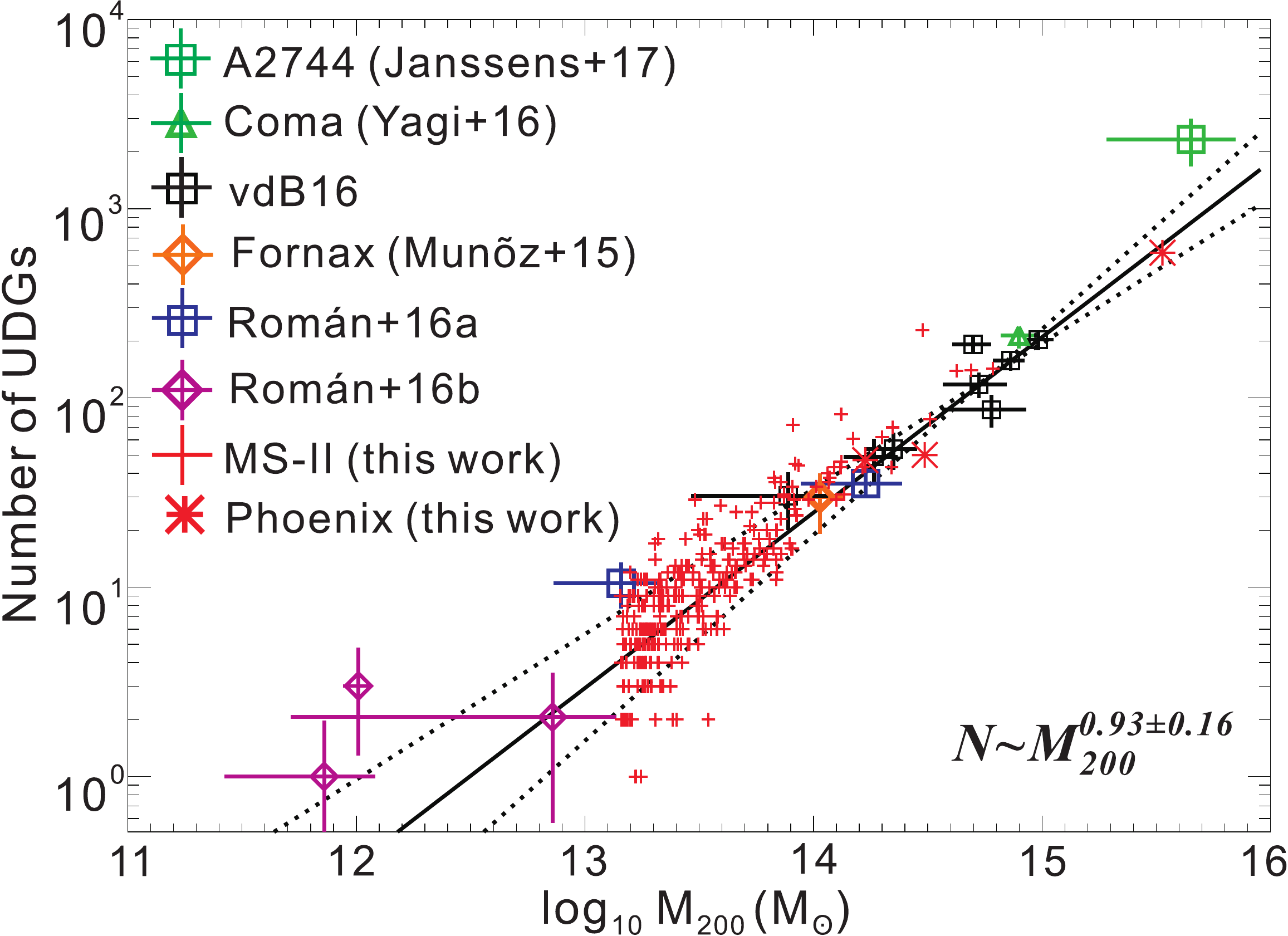}
\caption{Abundance of UDGs as a function of their host cluster
  mass. Solid and dashed lines reveal the relation $N\propto  
  M_{200}^{0.93\pm 0.16}$ obtained by Janssens et al. (2017). The red
  crosses and stars show the model predictions in
  MS-\uppercase\expandafter{\romannumeral2} and Phoenix simulations respectively,
  while the other colored symbols show the abundances of the observed
  UDGs in clusters and groups.}
\label{abundance}
\end{figure}

The abundances of UDGs are observed to correlate with the mass of their
host cluster \citep{Janssens17,vanderBurg16}: $N\propto
M_{200}^{0.93\pm0.16}$, where $N$ is the number of UDGs in a cluster
and $M_{200}$ is the cluster mass within a radius, $r_{200}$, within
which the average density is 200 times the cosmic critical density. In
Fig.~\ref{abundance}, we show the observed results, as well as our
model predictions of the abundances of UDGs as a function of their host
cluster masses. In order to compare with the observations directly,
here we discard the model UDGs fainter than $\mu_{0,g}\sim
26.5\ \rm{mag/arcsec^2}$ (approximately corresponding to the $r$-band
$ \langle\mu\rangle_{{\rm{e}},r} \leq 27\ \rm{mag/arcsec^2}$). It
shows that the model predictions are in excellent agreement with the
observed abundance-mass relation, from groups to rich clusters.

{\bf Surface number density (SND) profile}
\begin{figure}
\centering
\includegraphics[scale=0.45]{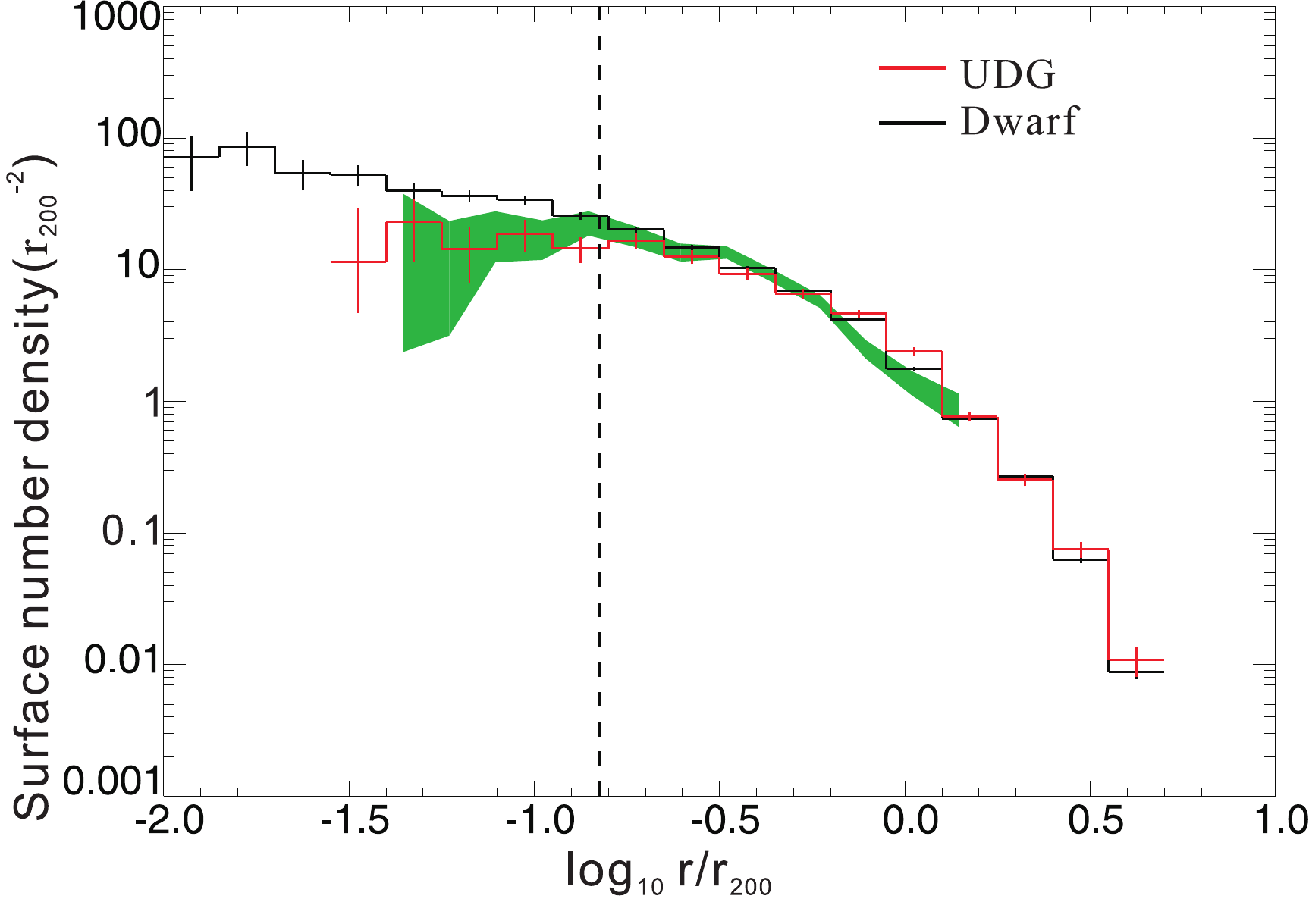}
\caption{Surface number densities as a function of $r/r_{\rm{200}}$
  for the model UDGs (red) and dwarf counterparts (black),
  respectively. The vertical dashed line denotes
  $r/r_{\rm{200}}=0.15$. The green component is the rescaled observed
  SND obtained by van der Burg et al. (2016).}
\label{snd}
\end{figure}

Spatial distribution of the galaxies in clusters provides important
clues to their evolution. In Fig.~\ref{snd}, we compare the SND
profile of the observed UDGs in 8 observed clusters
\citep{vanderBurg16} with the average SND profile of the model UDGs in
the 10 simulated clusters ($M_{200}>10^{14}\ M_{\odot}/h$) selected
from the MS-\uppercase\expandafter{\romannumeral2} simulation. The
observed SND profile is rescaled by a constant factor to account for
the different normalization methods. Clearly, the predicted SND
profile fits very well with the observed one over all the observed
scales from 0.03$r_{200}$ to $r_{200}$, including the flatting feature
in the inner part.

For completeness, we also show the SND profile of the typical dwarf
counterparts (the black histogram) in Fig.~\ref{snd}. The SND of UDGs
is similar to the profile of the dwarfs at $r/r_{\rm{200}}>0.15$
($r/r_{\rm{200}}=0.15$ is denoted by the vertical dashed line in
Fig.~\ref{snd}), while it is significantly lower at
$r/r_{\rm{200}}<0.15$. Moreover, UDGs are absent in the innermost
region $r/r_{\rm{200}}<0.03$. The lack of UDGs in the inner regions of
clusters could be caused by two possible reasons:
(\expandafter{\romannumeral1}) UDGs might have been disrupted and
dissociated by the strong tidal forces in the inner regions;
(\expandafter{\romannumeral2}) UDGs might have fallen into the clusters
more recently than the dwarf counterparts so that they have not
arrived in the inner regions yet. Observationally, the evidence of
tidal disruption for UDGs is very rare \citep{Mihos15,Toloba16}. We
will show in section~4 that our model indeed supports the second
explanation.


{\bf Color}

\begin{figure}
\centering
\includegraphics[scale=0.35]{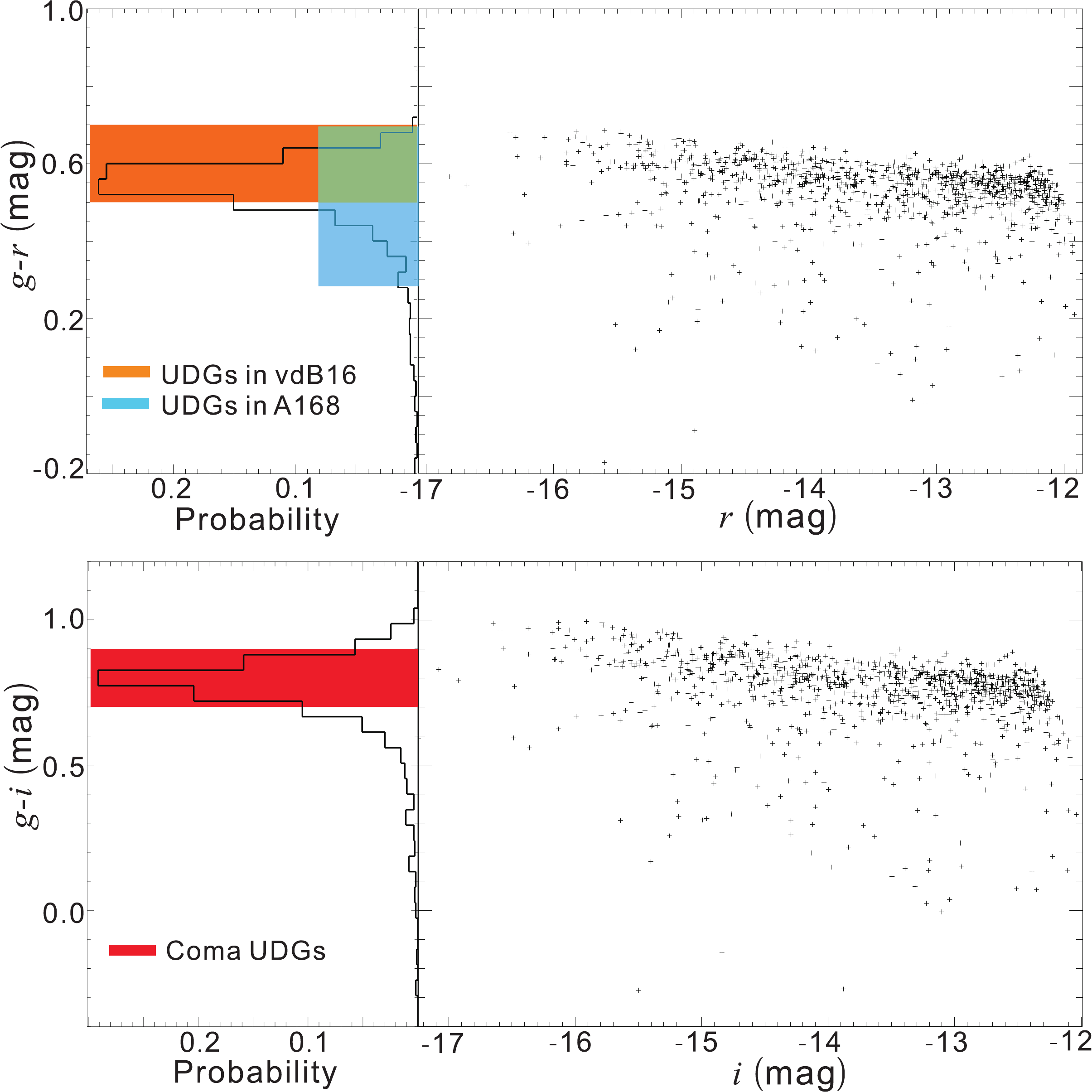}
\caption{Distributions of colors and color vs. magnitude diagrams of
  the model UDGs in the clusters. The upper and lower panels show the
  $g-r$ color versus the absolute $r$-band magnitudes, and $g-i$ color
  versus $i$, respectively. The orange, blue, and red components
  represent the approximate color ranges of the observed UDGs in the 8
  low-redshift clusters (van der Burg et al. 2016), A168 (Rom\'an et
  al. 2016a), and Coma (van Dokkum et al. 2015a), respectively.}
\label{cmd}
\end{figure}

Previous work found that except for several UDGs in groups
\citep[e.g.,][]{Roman16a}, most of the observed UDGs are red
\citep[e.g.,][]{vanDokkum15a,Koda15,vanderBurg16}. Fig.~\ref{cmd}
displays the color-magnitude diagrams for the model UDGs in our
simulated clusters. Analogous to the observations, most of the model
UDGs are red, except for several relatively faint ones. The left
panels show the color distributions of the model UDGs (black
histograms) and the 1$\sigma$ range  of the observed ones 
(colored regions) in clusters \citep{vanderBurg16, vanDokkum15a}. Most
of the model UDGs are located at $g-r\sim 0.6\pm 0.1$ and $g-i\sim
0.8\pm 0.1$, in good agreement with the observations.

{\bf Morphology} 

\begin{figure}
\centering
\includegraphics[scale=0.45]{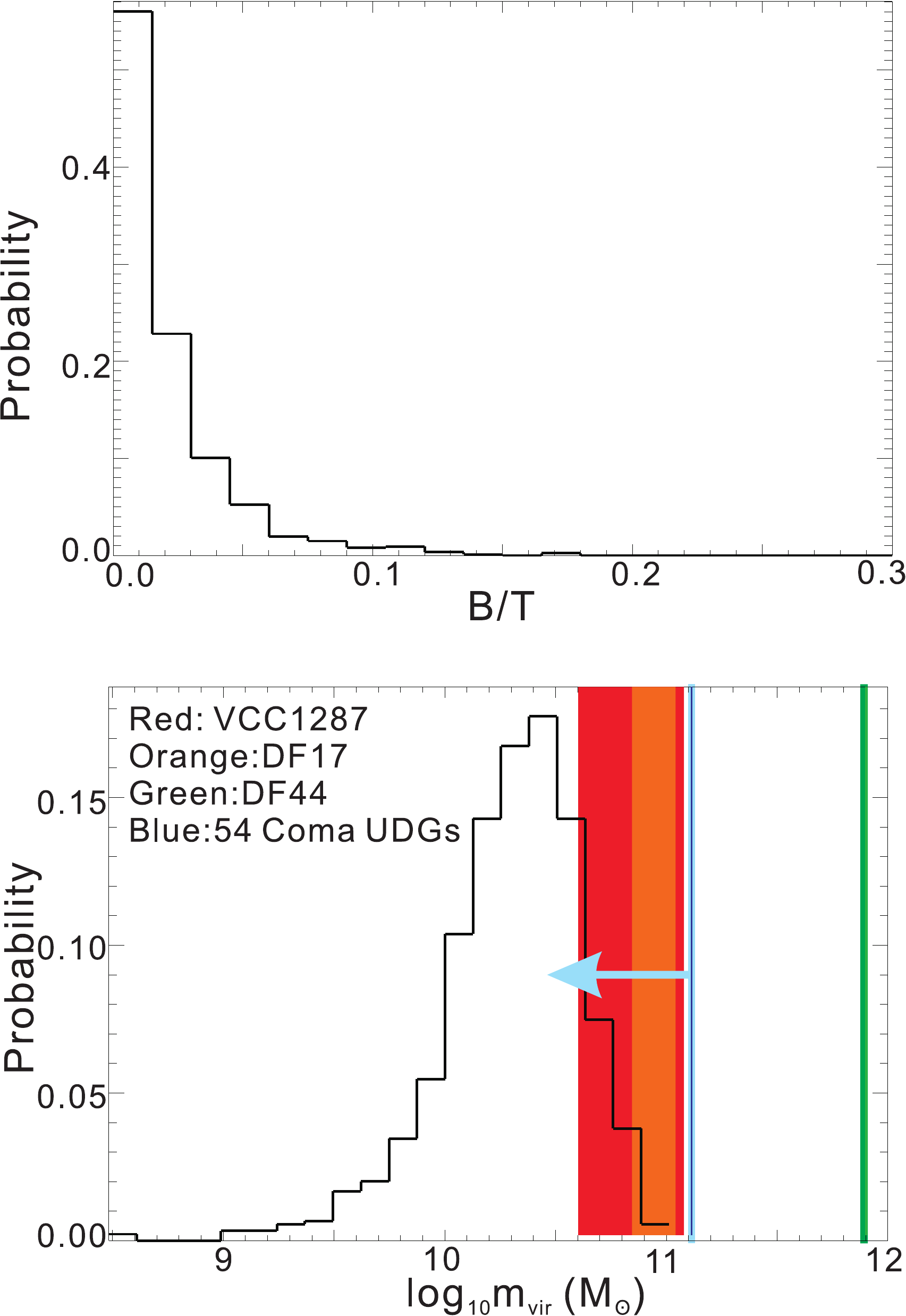}
\caption{Distributions of $B/T$ (the upper panel) and $m_{\rm{vir}}$
  (the lower panel) for model UDGs in the clusters, respectively. The
  colored components denote the virial masses and their errors of some
  observed UDGs. The arrow indicates upper limit of that the
  virial masses of the 54 Coma UDG candidates.}
\label{comparison}
\end{figure}

Observationally, UDGs are found to have low S\'ersic indices, $n\sim
0.6\--1$ \citep[e.g.,][]{vanDokkum15a,vanderBurg16,Munoz15,Koda15}. Limited
by the capability of semi-analytic models, we cannot measure the
profiles directly. Instead here we use the bulge-to-total mass ratio,
$B/T$ \citep{Weinzirl09}, as a proxy, i.e., a lower S\'ersic index
corresponds to a lower value of $B/T$. The upper panel of
Fig.~\ref{comparison} shows that most (95.7\%) of the model UDGs have
extremely low $B/T$ ($B/T<0.1$), in line with the observed low S\'ersic indices.

Most of our model UDGs present $B/T<0.1$, suggesting that the model UDGs perhaps have an oblate, disk-like geometry. This result may conflict with the conclusion of \cite{Burkert16}, who claimed that UDGs are more likely to be prolate rather than oblate because of their observed axial ratio range $q\sim 0.4\--1.0$. However, this conflict may be due to the fact that most of the observed UDGs have relatively small inclination angles $\theta$, i.e., they are more likely to be ``face-on'' rather than ``edge-on''. This is because the UDGs with large inclination angles may be too bright to be identified as `UDGs'. For instance, for a UDG with $\langle\mu\rangle_{\rm{e}}\sim 26\ \rm{mag/arcsec^2}$ and $\theta \sim 70^{\circ}$, its projected area decreases by a factor of $(\cos\theta)^2$, implying that the surface brightness changes about $2.5\log_{10}(\cos\theta)^2\simeq 2.3$~mag; therefore its apparent surface brightness becomes $\langle\mu\rangle_{\rm{e}}\sim 23.7\ \rm{mag/arcsec^2}$, which would be beyond the criterion to select UDGs.

{\bf Total mass}

The observed UDGs are very diffuse, and can reside in the dense environments without significant evidence of tidal disruption, suggesting that they are highly dark-matter dominated systems. Observationally, we usually use the abundance of the member globular clusters to infer UDG virial mass, $m_{\rm{vir}}$. In the lower panel of Fig.~\ref{comparison}, we show the distribution of the virial masses of the model UDGs. Most of them are in the range of $ 10^9\-- 10^{11}\ M_{\odot}$, consistent with the recently reported total masses for VCC~1287, DF17, and UGC2162 \citep{Beasley16,BeasleyTrujillo16,Trujillo17}, and Fornax UDGs \citep{Zaritsky16}, as well as 54 Coma UDGs \citep{Amorisco16b}. The peak of the model predictions is lower than the observations, which is primarily because that the measured UDGs are observationally brighter.  
Note that one particular case, DF44, is reported (van Dokkum et al. 2016) to be hosted in a dark halo as massive as $m_{\rm{vir}}\sim 8\times 10^{11}\ M_{\odot}$, an order of magnitude more massive than the typical mass found for the other UDG hosts. \cite{Zaritsky16} argued that DF44 lies at the upper-end in the size-enclosed mass relation of the observed UDGs and thus may not be a typical UDG. 


\subsection{UDGs in the Local Group and Local Volume}

\begin{figure}
\centering
\includegraphics[scale=0.33]{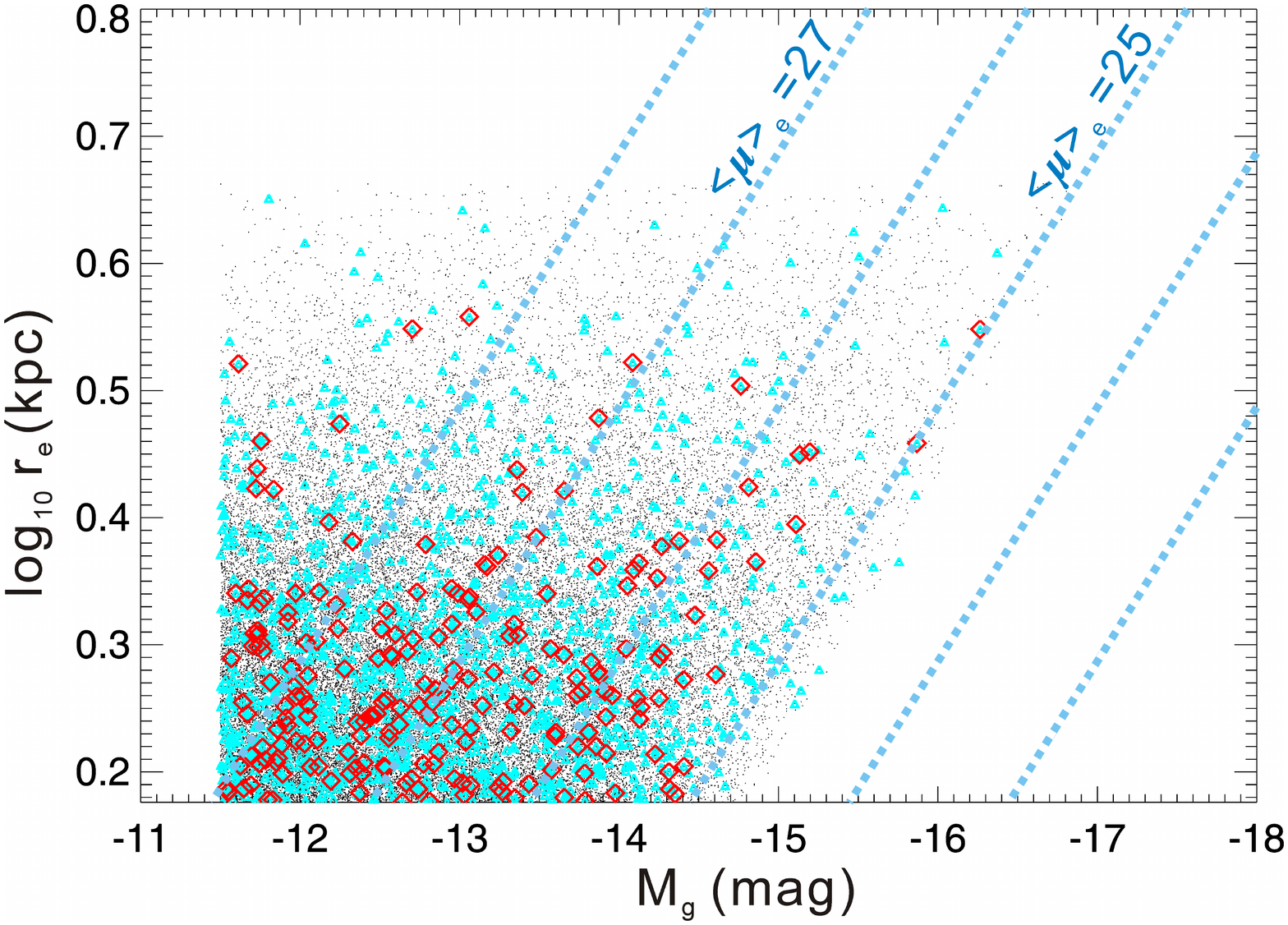}
\caption{$r_{\rm{e}}\-- M_g$ relation for UDGs in the Local Group and
  in the Local Volume. Black points denote all of the model UDGs
  selected in section 2.3. Red diamonds and cyan triangles denote the
  UDGs in the Local Groups and the Local Volumes analogue,
  respectively. Analogous to Fig.~\ref{re_mag}, the blue dotted lines
  highlight $\langle\mu\rangle_{\rm{e}}\simeq 23\sim
  28\ \rm{mag/arcsec^2}$.}
\label{z0_local}
\end{figure}

Most UDGs are discovered in rich clusters and their surrounding. A
very interesting issue is whether UDGs could exist in the Local Group
($\sim 5\times 10^{12}\ M_{\odot}$, e.g. Li \& White 2008).
In the literature, two UDG candidates are found in the census of faint
galaxies in the Local Group: Sagittarius dSph \citep{McConnachie12}
and Andromeda
\uppercase\expandafter{\romannumeral10}\uppercase\expandafter{\romannumeral10}\uppercase\expandafter{\romannumeral10}\uppercase\expandafter{\romannumeral2}
\citep{Martin13}. Sagittarius dSph is 26~kpc away from us, with
$V$-band absolute magnitude of $M_V=-13.5$~mag, $r_{\rm{e}}=2.6$~kpc,
and $V$-band $\mu_0\simeq 25.2\ \rm{mag/arcsec^2}$. This UDG candidate has been reported by \cite{Yagi16}, who are the first authors to identify UDGs in the Local Group. Andromeda
\uppercase\expandafter{\romannumeral10}\uppercase\expandafter{\romannumeral10}\uppercase\expandafter{\romannumeral10}\uppercase\expandafter{\romannumeral2}
is 0.78~Mpc away from us, with $M_V=-12.3$, $r_{\rm{e}}=1.46$~kpc
(slightly smaller than 1.5~kpc), and $\mu_0=26.4\ \rm{mag/arcsec^2}$.

To compare the model predictions with the data, we first define the
Local Group analogues in the simulation according to the observable
properties of the Local Group. Here we adopt the selection criteria
similar to those described in \cite{Xie14}. We first select the Milky
Way analogue using the criteria: $B/T<0.5$ and $5.4\times
10^{10}<M_{\rm{MW}}<7.4\times 10^{10}\ M_{\odot}$, where $M_{\rm{MW}}$
is the stellar mass of the Milky Way analogue, and then request that
there is only one bright companion (M31 analogue) within 1~Mpc from
each Milky Way analogue, with stellar mass
$M_{\rm{MW}}<m_{\rm{st}}<2M_{\rm{MW}}$ (this mass restriction is
slightly different from the criterion used in Xie et al. 2014). We
further require no galaxy clusters with masses $\geq
10^{14}\ M_{\odot}$ within 10~Mpc of the Local Group catalog. In
total, we find 69 ``Local Group'' analogues in the model galaxy
catalog. 207 model UDGs are found within 1~Mpc of the 69  ``Local
Groups'' analogue, i.e. 3 UDGs in each system on average. The
$r_{\rm{e}}$ vs. $M_g$ relation of these 207 UDGs are overplotted with
the red diamonds in Fig.~\ref{z0_local}.

We further extend the searching radius from 1~Mpc to 5~Mpc as the
Local Volume analogue. There are in total 1654 model UDGs (cyan
triangles in Fig.~\ref{z0_local}) in the simulated Local Volumes,
corresponding to 24 UDGs in each system. Observationally, we use the
dwarf catalog by
\cite{Karachentsev13}{\footnote{http://www.sao.ru/lv/lvgdb/}} to
search for the UDG candidates and find 23 possible UDGs residing in the
Local Volume, as listed in Table.~\ref{UDG_LV}. Note that these
galaxies are observed in a different wavelength and their $\mu_0$ are not given by \cite{Karachentsev13}, 
we thus use the selection criteria slightly different from those in Sec 2.3 by
requiring: a linear Holmberg diameter $A_{26}>3$~kpc,
$\langle\mu\rangle_B\geq 25\ \rm{mag/arcsec^2}$, and $10^6<L_K<10^9\ L_{\odot}$. Among
these possible UDG candidates, CenA-MM-Dw3 has been reported
\citep{Crnojevic16}.

In summary, the predicted abundances of UDGs in the simulated Local Group
analogue and Local Volume analogue agree very well with those in the
real Universe. In addition, we find that the model UDGs comprise less
than 10\% of the total faint populations ($M_g\sim $-17 $\--$
-11.5~mag) in these two systems and thus will not significantly affect
the corresponding conditional luminosity functions by
including/excluding UDGs.

\begin{table} \small
\begin{tabular}{@{}lccccc@{}}
\hline
\hline
Name & D & $A_{26}$ & $M_B$ & $\langle\mu\rangle_B$ & $\log L_K$ \\
Col. (1) & Col. (2) & Col. (3) & Col. (4) & Col. (5) & Col. (6) \\
\hline
Sag dSph & 0.02 & 3.08 & -12.67 & 26.08 & 8.02\\
And \uppercase\expandafter{\romannumeral10}\uppercase\expandafter{\romannumeral10}\uppercase\expandafter{\romannumeral10}\uppercase\expandafter{\romannumeral2} & 0.78 & 3.60 & -11.53 & 27.56 & 7.56 \\
NGC3109 & 1.34 & 7.73 & -15.75 & 25.01 & 8.58 \\ 
DDO099 & 2.65 & 3.24 & -13.53 & 25.34 & 7.42 \\ 
KK35 & 3.16 & 3.91 & -14.30 & 25.00 & 7.97 \\
KKH12 & 3.48 & 3.44 & -13.35 & 25.65 & 7.80 \\
MB3 & 3.48 & 5.19 & -13.97 & 25.92 & 8.22 \\
Cam A & 3.56 & 4.61 & -13.85 & 25.78 & 7.79 \\
CenA-MM-Dw1 & 3.63 & 3.13 & -12.56 & 26.23 & 7.98 \\
IKN & 3.75 & 3.15 & -11.63 & 27.17 & 7.60 \\
ESO269-058 & 3.75 & 5.52 & -15.04 & 25.00 & 8.86 \\
KK77 & 3.80 & 3.15 & -12.22 & 26.58 & 7.84 \\
HolmIX & 3.85 & 3.15 & -13.75 & 25.06 & 7.75 \\
HolmI & 4.02 & 5.54 & -14.59 & 25.44 & 8.05 \\
LV J1228+4358 & 4.07 & 4.56 & -13.94 & 25.67 & 7.83 \\
UGC A442 & 4.37 & 7.52 & -14.71 & 25.98 & 8.03 \\
DDO169 & 4.41 & 3.65 & -13.80 & 25.32 & 7.73 \\
IC3687 & 4.57 & 6.92 & -14.60 & 25.91 & 8.19 \\
CenA-MM-Dw3 & 4.61 & 6.63 & -12.32 & 28.10 & 7.88 \\
DDO226 & 4.92 & 3.12 & -13.63 & 25.15 & 7.71 \\
DDO126 & 4.97 & 4.14 & -14.42 & 25.00 & 8.09 \\
KK208 & 5.01 & 8.77 & -14.39 & 26.64 & 8.71 \\
ESO115-021 & 5.08 & 10.14 & -15.58 & 25.76 & 8.75 \\
\hline
\hline
\end{tabular}
\caption{Parameters of the 23 UDG candidates selected from the dwarf catalog of Karachentsev et al. (2013). Col. (1): Dwarf Name. Col. (2): Distance (Mpc) to the Milky Way. Col. (3): The linear Holmberg diameter in unit of kpc (Karachentsev et al. 2004; Karachentsev et al. 2013). Col. (4): $B$-band absolute magnitude. Col. (5): Mean surface brightness. Col. (6): Logarithm of $K$-band luminosity ($L_{\odot}$).}
\label{UDG_LV}
\end{table}


\section{Distribution of the UDGs in the Universe}
In the last section, we demonstrated that the model reproduces most of
the available observational properties of UDGs. This encourages us
to use our model to make prediction of the distribution of UDGs in the
Universe.

\begin{figure}
\centering
\includegraphics[scale=0.3]{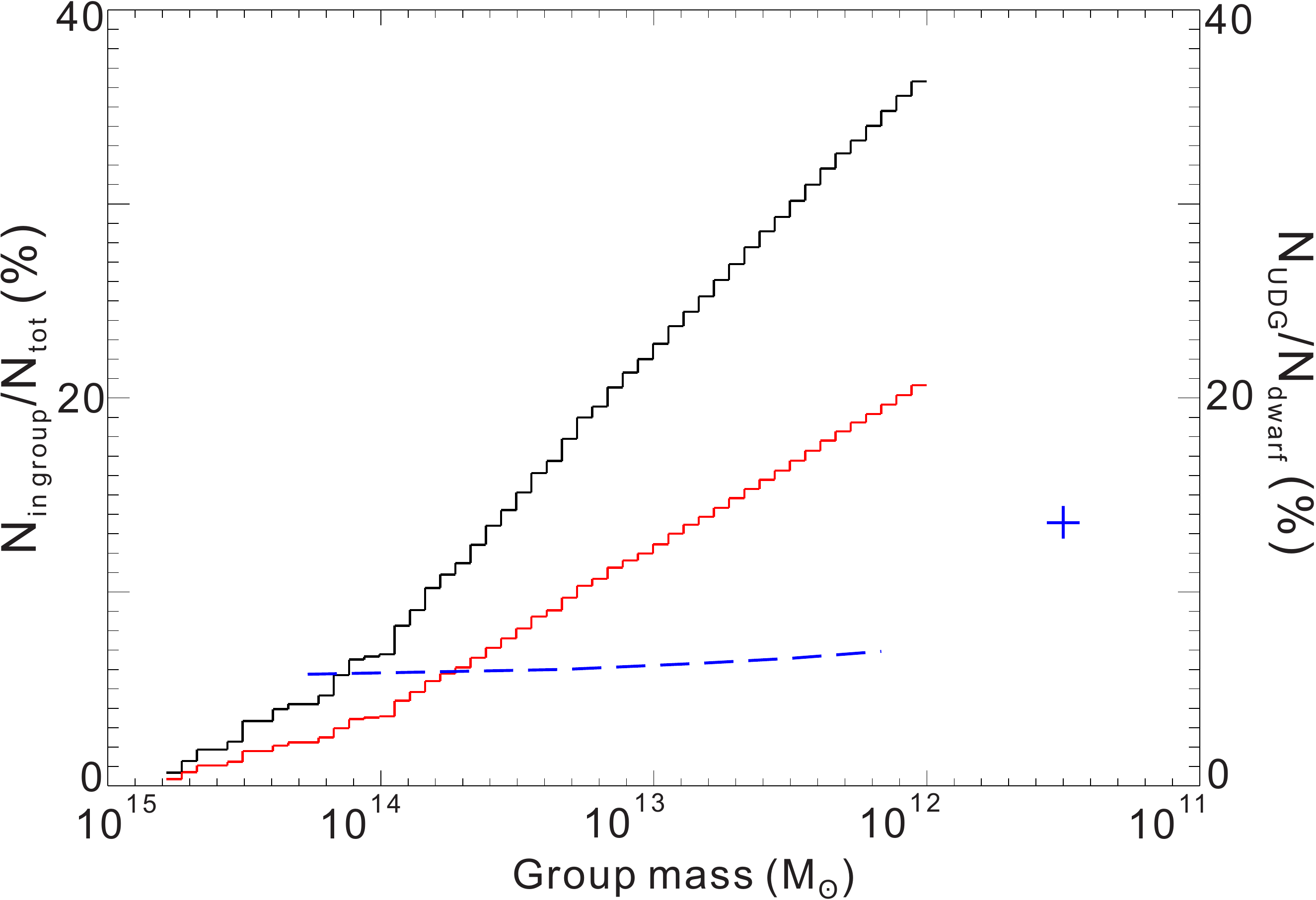}
\caption{Solid histograms show the cumulative fractions of the model
  UDGs (red) and dwarfs (black) in groups with the different masses
  (left $y$-axis). Dashed curve shows the ratio between the abundances
  of the model UDGs and dwarfs as a function of the host group mass
  (right $y$-axis). The blue cross marks the ratio in the fields.}
\label{fig:UDG_Envi}
\end{figure} 

Fig.~\ref{fig:UDG_Envi} shows the accumulative fraction of the model
UDGs as a function of the group masses. We find that only 4\% model
UDGs reside in the clusters more massive than $10^{14}\ M_{\odot}$, while
most of the model UDGs (80\%) reside in the groups with
$M_{200}<10^{12}\ M_{\odot}$ or in the fields. We further find that
most of the model UDGs are red in color regardless of their
environments, yet with an expected positive correlation between the
red fractions and densities of environments.

Comparing the distributions of the model dwarf galaxies and UDGs, in the
clusters and groups (more massive than $10^{12}\ M_{\odot}$), 7\% of
the dwarf galaxies are classified as UDGs, and the fraction is nearly
independent of the host halo mass. However in the fields, this
fraction is as high as 14\%, suggesting that UDGs tend to stay in the
lower-density environment. The general low fractions of UDGs also suggest that the luminosity function at the faint end will not be significantly
affected by including/excluding UDGs; the effect of UDGs on the
conditional luminosity functions is even weaker in the
groups and clusters.

\begin{figure*}
\centering
\includegraphics[scale=0.45]{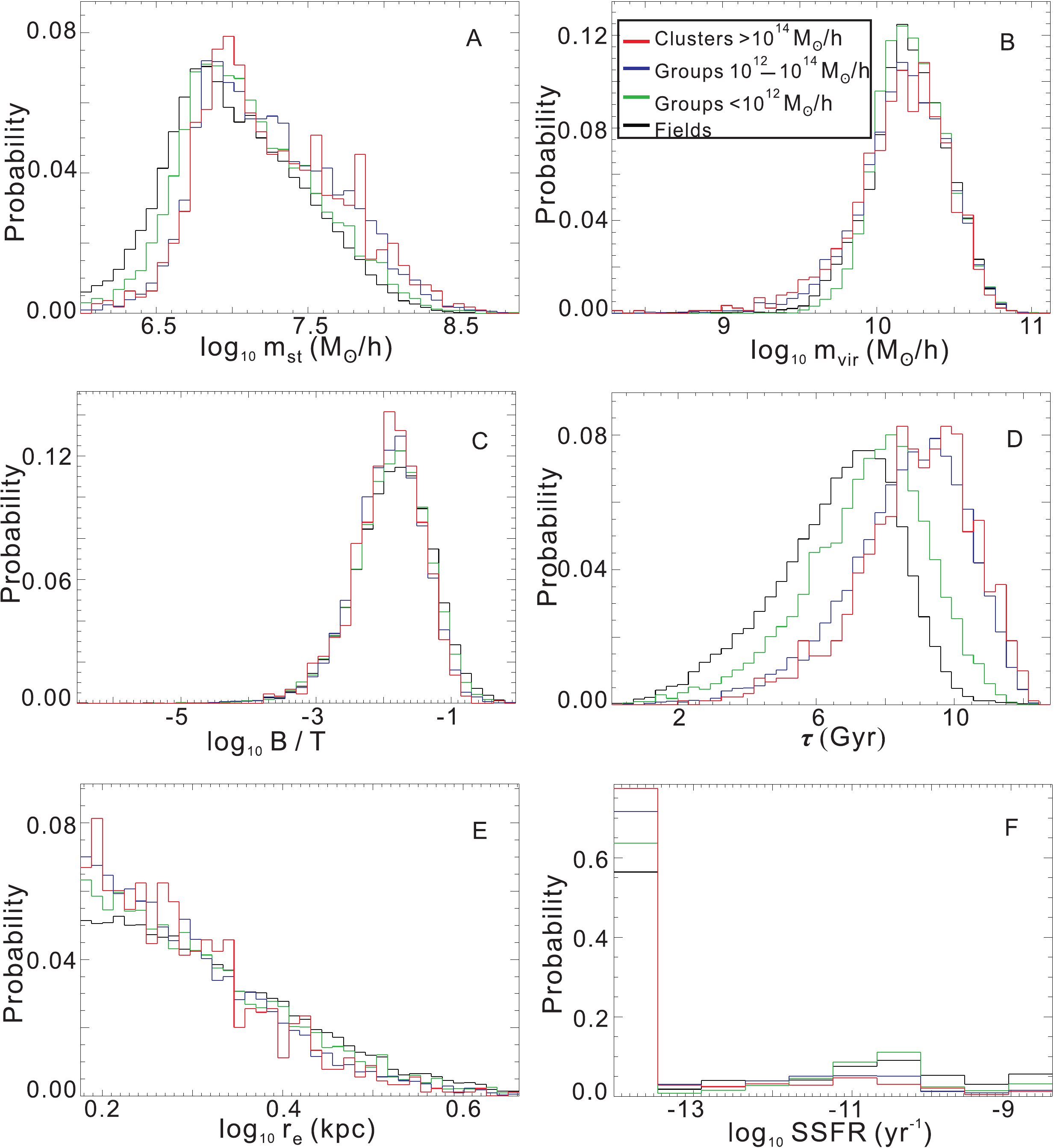}
\caption{A - F panels show the distributions of $m_{\rm{st}}$,
  $m_{\rm{vir}}$, $B/T$, $\tau$, $r_{\rm{e}}$, and SSFR,
  respectively. The red, blue, green, and black histograms denote the
  UDGs in clusters ($m\geq 10^{14}\ M_{\odot}/h$), in groups ($m\sim
  10^{12}\-- 10^{14}\ M_{\odot}/h$), in galaxy systems
  ($m<10^{12}\ M_{\odot}/h$), and in fields, respectively.}
\label{UDG_region_z0}
\end{figure*}

In order to study the environmental dependence of the model UDGs in
more details, we further divide the model UDGs into four subsamples
according to their host halo masses, clusters ($m\geq
10^{14}\ M_{\odot}/h$), groups ($m\sim 10^{12}\--
10^{14}\ M_{\odot}/h$), galaxy systems ($m< 10^{12}\ M_{\odot}/h$),
and fields. UDGs in the former three systems exist as satellite
galaxies, while in the fields, they are central galaxies of their own
halos (i.e., isolated UDGs). In Fig.~\ref{UDG_region_z0}, we show the
probability distributions of six different physical properties for the 4
UDG subsamples. The stellar masses, $m_{\rm{st}}$, of the model UDGs
increase with the densities of environments (panel A), e.g.,
$m_{\rm{st}}$ of the UDGs in the clusters is higher by about
0.2~dex than those in the fields. This is consistent with the
increasing fraction of the red populations with the increasing
environmental density, as the red galaxies are usually older and more massive (for a given luminosity). UDGs in clusters tend to have lower specific star-formation rates (SSFR=star-formation rate/$m_{\rm{st}}$; panel F) and be relatively older (panel D). Galaxies formed earlier (with
higher mass-weighted-ages, $\tau$) are usually more compact, which is
reflected by the distributions of $r_e$ (panel E). Different from the
other properties, dependences of the virial mass $m_{\rm{vir}}$
(panel B) and morphology (panel C) on the environments are very
weak. Regardless of the environments, most of the model UDGs are
disk-dominated systems and formed in the halos of virial mass $\sim
10^{10}M_{\odot}$, very similar to that of a typical dwarf galaxy.

\section{Formation of UDGs}

As discussed above, our model predicts that UDGs have the similar dark
matter halos to those of the typical dwarf galaxies. In this section, we
explore why the stellar components of UDGs are so extended.

\begin{figure*}
\centering
\includegraphics[scale=0.45]{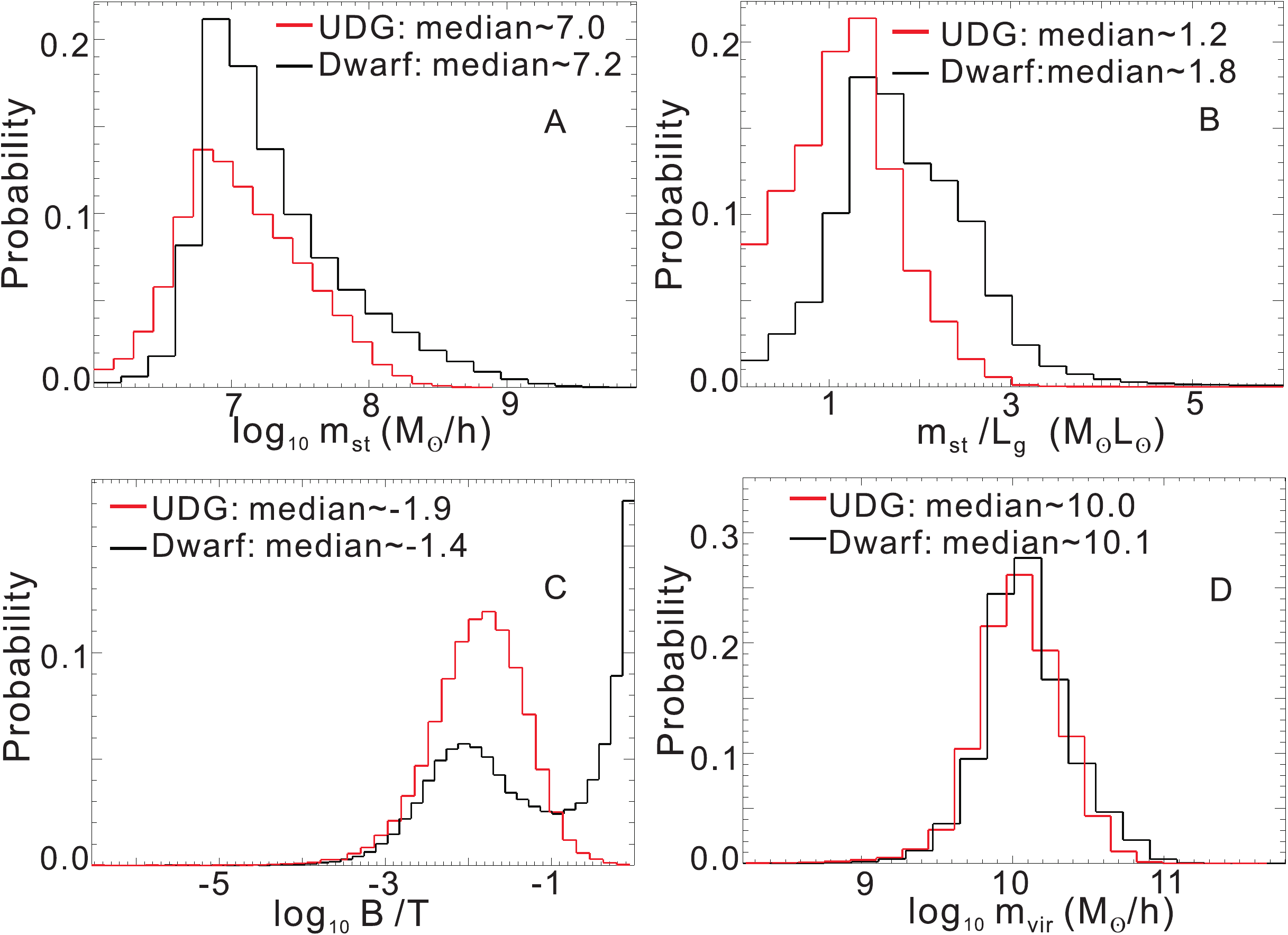}
\caption{A - D panels show the distributions of $m_{\rm{st}}$,
  $m_{\rm{st}}/L_g$, $B/T$ and $m_{\rm{vir}}$, respectively. Red and
  black histograms represent the distribution of UDGs and their dwarf
  counterparts, respectively. The median values of each quantity are
  shown in the corresponding panel.}
\label{UDG_dwarf_z0}
\end{figure*}

\begin{figure}
\centering
\includegraphics[scale=0.4]{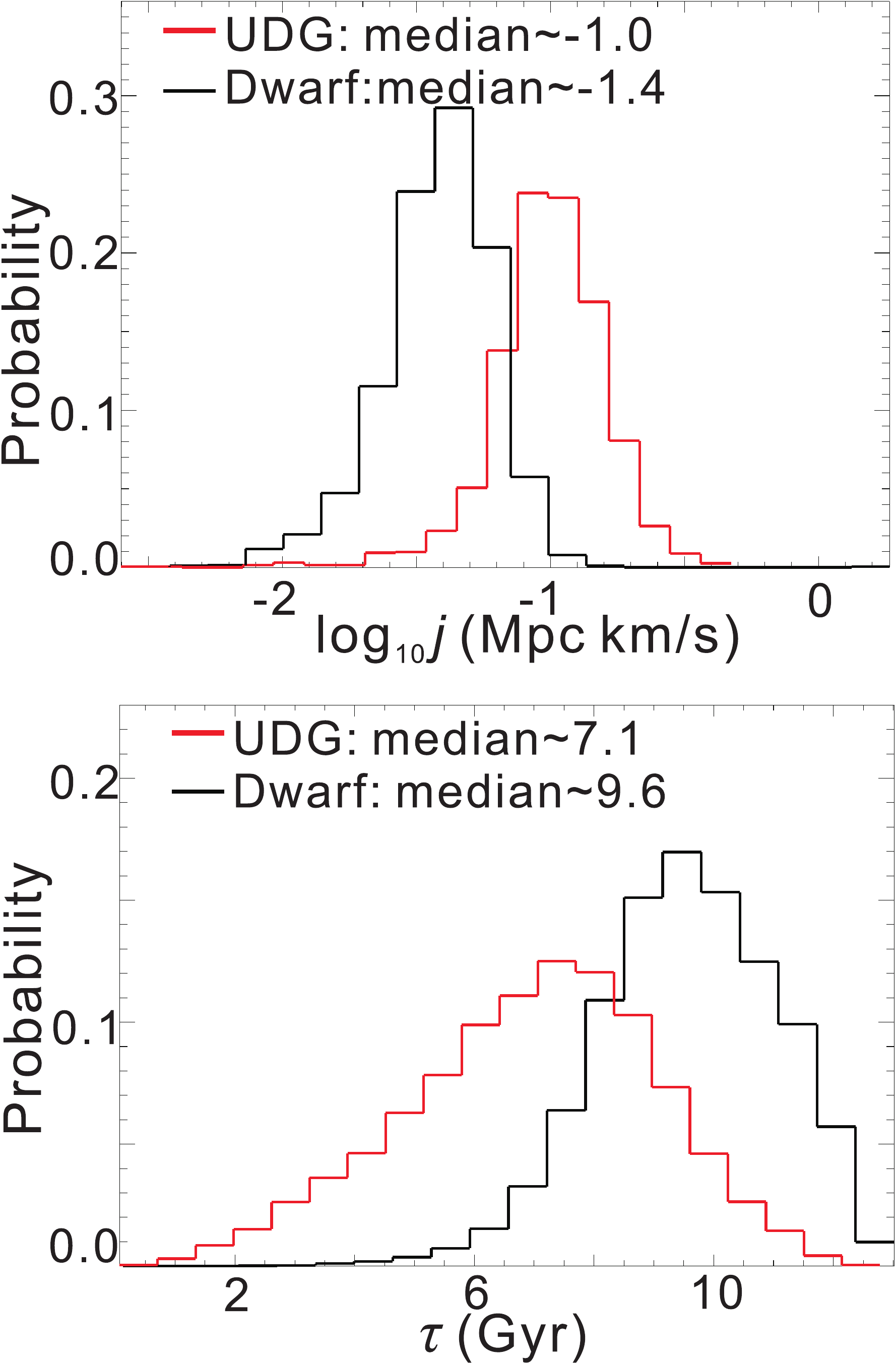}
\caption{The distributions of the specific angular momenta $j$ of the progenitor halos (upper panel) at $t_{\rm{half}}$ and galactic ages
  $\tau$ (lower panel) for the model UDGs (red histogram) and dwarf counterparts
  (black histogram), respectively.}
\label{UDG_dwarf_zf}
\end{figure}

In order to investigate the differences between the model UDGs and
typical dwarf galaxies more clearly, in Fig.~\ref{UDG_dwarf_z0}
we compare the distributions of the stellar mass (panel A), stellar
mass-to-light ratio (panel B), $B/T$ (panel C) and host halo mass
(panel D) between the two samples. The typical dwarfs in general
have larger stellar masses than UDGs (panel A). This is primarily
because that the model UDGs are younger (see the
lower panel in Fig.\ref{UDG_dwarf_zf}) and bluer. Consistently, the stellar
mass-to-light ratio of a UDG is typically smaller than that of a
typical dwarf galaxy by around 50\% (panel B). One needs to pay
particular attention when converting the luminosities of UDGs to their
stellar masses. UDGs are much more extended than the regular dwarf
galaxies, and thus we expect that the internal structures of UDGs
differ from the dwarf counterparts as well. As shown in the panel C, $B/T$ of the model UDGs are significantly lower than those of the
typical dwarf counterparts. 96.6\% of the model UDGs present $B/T<0.1$;
whereas about 27\% of the dwarf counterparts are dEs
with $B/T>0.5$. Regardless of the differences shown above, the
distributions of the virial masses, $m_{\rm{vir}}$, of the model UDGs
and dwarf counterparts are very similar to each other. The host halos
of UDGs are only slightly less massive than those of their dwarf
counterparts by 0.1~dex.

There are two reasons which may account for the unique feature of
UDGs: 1) UDGs may form much later than the typical dwarfs, since the
objects formed later are usually more extended because of the diluted
Universe at a low redshift; 2) UDGs may have much higher spin parameters as
naively expected from the standard galaxy formation scenario in which the
galaxy size $r_{\rm{e}}\propto \lambda R_{\rm{vir}}$ \citep[e.g.,][]{Mo98, Amorisco16},
where $\lambda$ and $R_{\rm {vir}}$ are the spin parameter and virial
radius of the host halo, respectively. We will examine these below with our model. Note that in the modern galaxy formation models, e.g., Guo11 and Bower et al. (2010), the size of a present galaxy is not uniquely determined by the spin parameter of its host halo at $z=0$ or any specific redshift, rather it is a cumulative consequence of the angular momentum evolution of its parent halo and star formation. The size of a galaxy is largely determined by the rotational states of its host halo when the galactic star formation rate was high.
We use the specific angular momentum of the main progenitor of a halo at the epoch ($t_{\rm{half}}$) when half of its stellar mass was assembled to take into account this integral effect. Note, the exact choice of the redshift only changes our result quantitatively but not qualitatively.  

In Fig.~\ref{UDG_dwarf_zf}, we present the distributions of the specific angular momenta $j$ of the progenitor halos at $t_{\rm{half}}$ and galaxy ages $\tau$ for the model UDGs and dwarf counterparts. Clearly, compared with the typical dwarfs, the specific angular momenta of the model UDGs are larger by a factor of 2.5 at $t_{\rm{half}}$. Also as shown in the lower panel of Fig.~\ref{UDG_dwarf_zf}, the UDGs are indeed much younger with a median age of 7.1~Gyr, compared with the typical dwarfs which have a median age of 9.6~Gyr. Further, these suggest that it is indeed the combination of the late formation of UDGs and high-spins of the host halos that result in the large sizes of UDGs. Therefore, the
high-spin tail origin of UDGs proposed by \cite{Amorisco16} is not the
complete story to explain the formation of UDGs. Besides, we find that almost all of the model UDGs in the clusters fall in directly from the field.

Di Cintio et al. (2017) developed an alternative strong-outflow model to explain the extended sizes of UDGs. Although this model can also reproduce the broad color range and low Sersic indics of observed UDGs, their simulated UDGs do not live in particularly high-spin halos, which conflicts with both of our model prediction and the recent observational spin parameters of UDGs from the ALFALFA HI survey \citep{Leisman17}. Besides, their simulation contains about 40 galaxies with halo masses of $10^{10}\--10^{11}\ M_{\odot}$ \citep{Wang15}, among which 8\--21 (depending on the effective radius threshold $r_{\rm{e}}=1$~kpc or 2~kpc) of them are UDGs, suggesting a much higher fraction of UDGs than our model prediction; in the sense that the outflow model may overestimate the adundance of UDGs.

\begin{figure}
\centering
\includegraphics[scale=0.25]{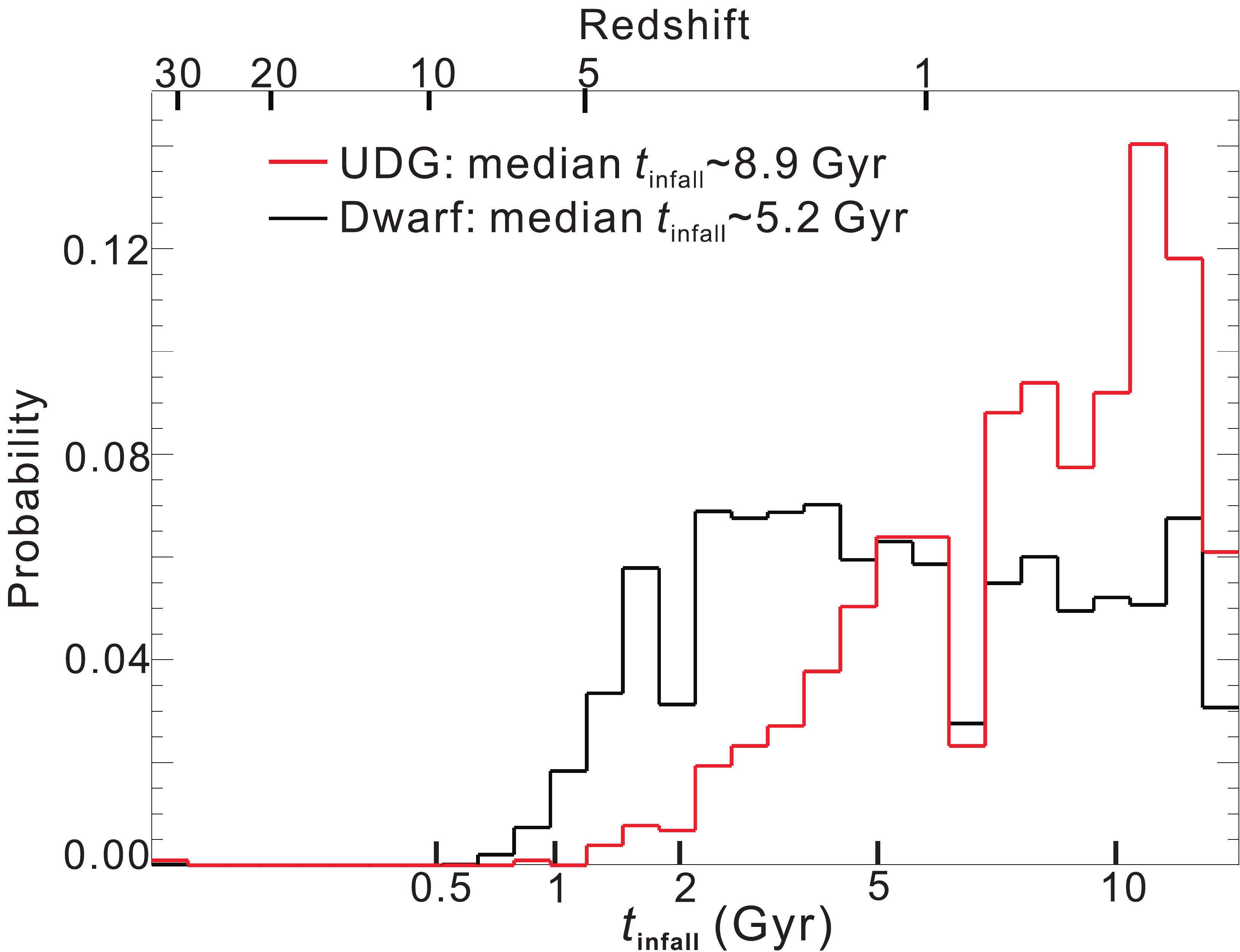}
\caption{Distributions of $t_{\rm{infall}}$ for the UDGs (red) and their dwarf counterparts (black) in the clusters, respectively.}
\label{age_dis}
\end{figure}

Another interesting phenomenon is that the number density profile of
UDGs is flat towards the center in the observed clusters, quite different
from that of the typical dwarfs. One possible reason may be that the
UDGs fell into the clusters later than the typical dwarfs. We examine
it in Fig.~\ref{age_dis} by comparing the distributions of the infall-time $t_{\rm{infall}}$ (the time at which a galaxy was accreted into a cluster; $t_{\rm{infall}}\simeq 13.75$~Gyr corresponding to $z=0$) of the two dwarf populations in the 10 simulated clusters. As expected
the infall-time of the model UDGs is on average significantly later 
than that of the dwarf counterparts, with a median value of $\langle
t_{\rm{infall}}\rangle\sim 8.9$~Gyr and $\langle
t_{\rm{infall}}\rangle\sim 5.2$~Gyr for the model UDGs and dwarf
counterparts, respectively. Therefore, the lack of UDGs in the inner regions of
clusters as shown in Fig.~\ref{snd}, as well as lack of the tidal
disruption features in observations are the natural consequences of this late
infall-time.

\section{Conclusion and Discussion}

As a special subset of low surface brightness population, UDGs draw
much attention recently because they are as faint as the typical
dwarf galaxies, yet have the sizes similar to those of the $L^*$
galaxies. The origin of UDGs is a mystery: are they the genuine dwarf
galaxies with extremely large sizes or failed $L^*$ galaxies? 

We use galaxy formation models (Guo et al. 2011, 2013) to study this
special galaxy population. The predicted properties of UDGs in the
clusters and groups (where most of the UDGs are discovered) agree very
well with the observational results, including the abundance, number
density profile, color distribution, and morphology, etc. Our model
predicts about 4 and 24 UDG candidates in the Local Group and Local Volume analogues, respectively. When searching for such candidates in the Local group and Local Volume with existing observational data, we find that the
numbers of UDGs in these two systems agree remarkably well with the
model predictions. 

We demonstrate that UDGs are genuine dwarf galaxies and can naturally
emerge from the $\Lambda$CDM model. It is the combination of the later
formation of UDGs and the relatively larger spins of their host halos
that results in the more extended feature of this particular
population. The lack of UDG candidates in the inner regions of
clusters and the lack of tidal disruption features can be naturally
explained by the later infall of the UDGs.

Compared to the typical dwarf galaxies, UDGs tend to reside in the
low density regions consistent with their later formation. However, in the fields where there is no environmental effect, UDGs are redder than the typical dwarf galaxies. This is
because the UDGs are more extended and the star formation ceases when
the densities of the gas disks drop below a certain threshold. The red
colors of UDGs suggest that it is even harder to detect UDGs than the
typical dwarf galaxies. Fortunately, the model predicts only 7\% of
dwarf galaxies in clusters and 14\% in fields are identified as UDGs,
suggesting that it will not significantly affect the global luminosity
function at the faint end, neither the conditional luminosity
functions. Although most of UDGs are discovered in dense environments,
we anticipate to discover a much higher fraction in under-dense regions in
the future.

\section*{Acknowledgments}
We thank Prof. Karachentsev I. D. for his helpful discussion on the
UDG candidates in the Local Volume. QG and LG acknowledges support
from NSFC grants (nos. 11425312), and two Newton Advanced Fellowships, as well as the hospitality of the Institute for Computational Cosmology at Durham University. THP acknowledges support in form of the FONDECYT Regular Project No. 1161817 and by the BASAL Center for Astrophysics and Associated Technologies (PFB-06).
\bibliographystyle{mn2e}


\end{document}